\documentclass[aps,prd,twocolumn,showpacs,superscriptaddress,amsmath,amssymb,letterpaper,floatfix]{revtex4}

\usepackage{dcolumn}
\usepackage{bm}
\usepackage{epsfig}
\usepackage{graphicx,psfrag}

\newcommand{\be}{\begin{equation}}
\newcommand{\ee}{\end{equation}}
\newcommand{\bear}{\begin{eqnarray}}
\newcommand{\eear}{\end{eqnarray}} \newcommand{\ba}{\begin{array}}
\newcommand{\ea}{\end{array}}

\def\beq{\begin{equation}}
\def\eeq{\end{equation}}
\def\beqa{\begin{eqnarray}}
\def\eeqan{\end{eqnarray}}

\def\to{\rightarrow}

\newcommand\iden{\leavevmode\hbox{\small1\normalsize\kern-.33em1}}

\def\W3{W_H^3}

\begin{document}
\setcounter{footnote}{0}

\title{Direct Detection of Kaluza-Klein Particles in Neutrino Telescopes}
\author{Ivone~F.~M.~Albuquerque}
\affiliation{Instituto de F\'isica, Universidade de S\~ao Paulo, 
S\~ao Paulo, Brazil}
\author{Gustavo Burdman}                     
\affiliation{Instituto de F\'isica, Universidade de S\~ao Paulo,
S\~ao Paulo, Brazil}
\author{Christopher A. Krenke}
\affiliation{Department of Physics, University of Arizona, Tucson, AZ
85721, USA}
\affiliation{Department of Physics, University of Maryland, College Park, 
MD 20742, USA}
\author{Baran Nosratpour}
\affiliation{Department of Physics, University of Arizona, Tucson, AZ 
85721,
USA}

\vspace*{0.3cm}


\begin{abstract}

In theories with universal extra dimensions (UEDs), all standard model 
fields propagate in the bulk and the lightest state of the first 
Kaluza-Klein (KK) level can be made stable by imposing a $Z_2$ parity. We 
consider a framework where the lightest KK particle (LKP) is a neutral, 
extremely weakly interacting particle such as the first KK excitation of 
the graviton, while the next-to-lightest KK particle (NLKP) is the first 
KK mode of a charged right-handed lepton. In such a scenario, due to its 
very small couplings to the LKP, the NLKP is long-lived. We investigate 
the production of these particles from the interaction of high energy 
neutrinos with nucleons in the Earth, and determine the rate of NLKP 
events in neutrino telescopes. Using the Waxman-Bahcall limit for the 
neutrino flux, we find that the rate can be as large as a few hundreds 
events a year for realistic values of the NLKP mass.

\end{abstract}
\maketitle

\section{Introduction} \label{intro} 

Although the standard model (SM) is a successful description of the energy 
scales experimentally probed so far, it is expected that new physics will 
appear at the TeV scale. This is precisely the energy regime soon to be 
studied at the Large Hadron Collider (LHC). It is also the natural scale 
for the dynamical origin of electroweak symmetry breaking, as well as for 
the solution of the hierarchy problem. Typical solutions of these problems 
involve either symmetries (e.g. supersymmetry), some dynamical mechanism 
of electroweak symmetry breaking (e.g. technicolor), or a combination of 
symmetry and dynamics (e.g. little Higgs). In a somewhat different class 
are extensions of the SM involving compact extra dimensions.  In Large 
Extra Dimensions \cite{led} only gravity propagates in the extra 
dimensional bulk, and the true fundamental scale of gravitation is 
$O(1)$~TeV. On the other hand, in theories with one curved extra 
dimension~\cite{rs1}, gravity is weak at the TeV scale due to the warping 
produced by the bulk metric.

Here we consider a more generic brand of extra dimensional theories, 
universal extra dimensions (UEDs), where all fields propagate in the 
extra dimensional bulk~\cite{bogdan}. Its main motivation is
phenomenological: if compact extra dimensions exist and all fields
propagate in them, the inverse compactification radius could be just
above the weak scale, setting the stage for a lot of new physics
possibilities at the TeV scale. Furthermore, adding a mild assumption,
the presence of a reflection symmetry leading to a $Z_2$-parity, UED
theories are endowed with a candidate for dark matter: the lightest KK
particle or LKP. 

Although at leading order the spectrum of each KK level is degenerate, it splits 
under radiative corrections, as well as when generic higher dimensional operators 
are taken into account~\cite{cms1}. In theories with one extra dimension, if only 
the loop contributions coming from the physics below the cutoff are considered, 
one obtains the spectrum of the minimal UED standard model (mUED) of 
Ref.~\cite{cms1}. In this case, the LKP is most likely to be the first KK mode of 
the photon $\gamma^{(1)}$. Other possibilities for the LKP include the first KK 
mode of the graviton $G^{(1)}$~\cite{FRT}, and (in theories where neutrino masses 
are Dirac) the first KK excitation of the right-handed neutrino 
$\tilde{N}^{(1)}$~\cite{MSSY}. Other light particles include the KK excitation of 
a right-handed charged lepton, $\ell^{(1)}$, and the charged Higgs KK 
mode~\cite{muedphase}. The splitting between the LKP and $\ell^{(1)}$ is 
typically only a few GeV, depending on the choice of parameters~\footnote{In UED 
theories with two extra dimensions~\cite{sixd}, the LKP is typically a neutral 
scalar adjoint, $B_H^{(1,0)}$. However, in principle the charged scalars 
$W_H^{(1,0)\pm}$ could be made lighter by higher dimensional operators, resulting 
in a similar situation as in 5D. We will not pursue this possibility here.}. 

The mUED spectrum is merely illustrative, and ultraviolet physics contributions 
to boundary terms could significantly alter it, making for instance, $\ell^{(1)}$ 
the NLKP, while either $G^{(1)}$ or $\tilde{N}^{(1)}$ remains the LKP.  In such a 
scenario, the decay of the NLKP to the LKP would be highly suppressed, making the 
NLKP lifetime very large. We will consider this possibility in this paper. This 
situation is analogous to what happens in some supersymmetric scenarios (e.g. 
gauge mediation) where the gravitino is the lightest supersymmetric particle and 
a right handed charged slepton is the next to lightest one. The phenomenology 
associated to a long-lived $\ell^{(1)}$ includes highly ionizing tracks at 
colliders. It also implies that $\ell^{(1)}$ can be produced by the interactions 
of high energy neutrinos with the earth and propagate through it until reaching a 
detector, in very close analogy to the case of NLSP sleptons studied in 
Refs.~\cite{abcprl, Bi, AKR, abcprd, AIMM}.

We will show that interactions of high energy neutrinos (E$_\nu>10^5$~GeV) 
with nucleons in the Earth will produce pairs of NLKPs. The rate of events 
will allow the discovery of the latter in km$^3$ neutrino telescopes. This 
analysis follows the same steps as for NLSP detection 
\cite{abcprl,abcprd}. The crucial observation is the same as for the NLSP, 
the small NLKP production cross section is compensated by its large range. 
The NLKP loses much less energy while traveling through the Earth when 
compared to SM leptons. This allows the detection of NLKPs that are 
produced far away from the detector.

As the NLKPs are produced in pairs, the main background consists of 
di-muon events. We will show that there are at least two ways to separate 
these from the signal. For lower mass NLKPs, the measured energy spectrum 
will have a bump in the region from $10^3$ to $10^4$ GeV due to the fact 
that the energy loss in the detector will resemble the one from lower 
energy muons. In addition, for both low or high mass NLKPs, the separation 
between the pair that crosses the detector will be larger than the one for 
the di-muon pair, and will allow to distinguish the signal from the 
background.

This paper is organized as follows: we first determine the NLKP production 
cross section; in Section~\ref{sec:dedx} we describe the NLKP energy loss 
while traveling through the Earth; the analysis of the signal and 
comparison with the background are discussed in Section~\ref{sec:signal} 
and the conclusions follow in the last section.

\section{NLKP production}
\label{sec:xsecs}
\begin{figure}
\centering
\epsfig{file=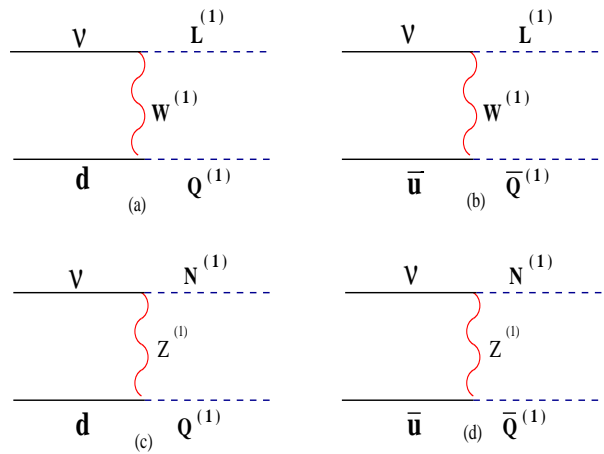,width=8cm,height=6cm,angle=0}
\caption{Feynman diagrams for KK mode production in $\nu N$ collisions.
Charged current (charged gauge boson KK mode) interactions: (a) Left-left interaction 
requiring the insertion of the gauge KK mode mass in the t-channel line. 
(b) Left-right interaction.  
Neutral current: (c), (d). 
There are analogous diagrams for anti-neutrinos as well as for 
strange and charm initial quarks. 
}
\label{fig:feynman}
\end{figure}

In this section, we compute the production cross section for the NLKP
pair. Due to the presence of the $Z_2$-parity, all KK modes produced
will eventually cascade down to a NLKP. Since KK modes are 
produced in pairs, 
KK production initiated 
by $\nu N$ scattering will result in a pair of NLKPs.   
The dominant process for $\nu-N$-initiated KK production involves the 
t-channel production of
a left handed lepton KK mode $L^{(1)}_i$ (with generation index $i$)
and a quark KK mode ($Q^{(1)}$) via a 
gauge boson KK mode ($W^{(1)}$). This process is analogous to the charged 
current (CC) in the SM. We also include the subdominant process which is analogous
to the neutral current process in the SM. This involves the exchange of a neutral
gauge boson KK mode ($Z^{(1)}$). These processes are shown in Figure~\ref{fig:feynman}.

The neutrino, which is always left-handed, can interact with a left handed down-type
quark (a) or with a right-handed up-type antiquark (b). 
This results in the partonic cross sections:
\beqa
\frac{d\sigma^{\rm (a)}}{dt} &=& \frac{8 \pi\alpha^2}{\sin^4\theta_W}\,
\frac{(s-m^2_{L^{(1)}_i}-m_{Q^{(1)}}^2)}{s^2 (t^2-M_{W^{(1)}}^2)^2}
\label{llcc} \\
\frac{d\sigma^{\rm (b)}}{dt} &=& \frac{8 \pi\alpha^2}{\sin^4\theta_W} \times \\ 
\nonumber 
 & & \frac{\left[m_{L^{(1)}_i}^2 \, m_{Q^{(1)}}^2 + u^2 - u\,(m_{L^{(1)}_i}^2 + m_{Q^{(1)}}^2)
\right]}{s^2(t^2 - M_{W^{(1)}}^2)^2} \\
\label{lrcc}
\eeqan
where $s,t$ and $u$ are the usual Mandelstam variables, and 
$m_{L^{(1)}_i},\,m_{Q^{(1)}}$ and $M_{W^{(1)}}$ are the $L^{(1)}_i,\,Q^{(1)}$ and
$W^{(1)}$ masses, repectively.  
The subdominant neutral gauge boson KK mode ($Z^{(1)}$) exchange 
is shown in Figure~\ref{fig:feynman}~(c)-(d). 
Each of these processes will produce a $L^{(1)}_i$ and a $Q^{(1)}$ and both of these
particles will promptly produce a decay chain ending with a
$\ell^{(1)}_i$.

Bounds from direct searches from the Tevatron, as well as from
electroweak precision 
constrains \cite{bogdan}, result in $R^{-1} >$  300~GeV for 5D, while
for 6D is $R^{-1} >$ 500 GeV. 
We will assume three illustrative values for the NLKP mass: 300, 600 and 900 GeV.
Finally, we need to specify the cutoff of the theory. 
Using naive dimensional analysis, we find for the 5D case, that 
$\Lambda R \sim 8 \pi / g^2$, where $g$ is the strong coupling constant.
We then take $\Lambda R\simeq 20$ as a sharp cutoff, i.e. we include
no contributions from energies above $20/R$. In order to evaluate the 
uncertainty introduced by this procedure in the cross section
calculation,  
we scanned values of $\Lambda R$ up to 30, 
with no significant effects in the results.

The NLKP production cross section is shown in Figure~2 as a function of
the neutrino energy. For comparison, the SM charged current (top gray curve) and the 
di-muon (solid red curve) background cross sections are also shown. As expected, the NLKP production
cross sections ($\sigma_{\rm NLKP}$) are significantly lower than the SM one. However, 
depending on the
neutrino energy and the $L^{(1)}_i$ mass, it can be larger than the di-muon
background. In the next section we will show that the the fact that
$\sigma_{\rm NLKP}$ is rather suppressed 
(as compared to the SM one) will be compensated by the sizable NLKP
range resulting from the combination of its long lifetime and small energy loss. 

\begin{figure}
\begin{center}
\epsfig{file=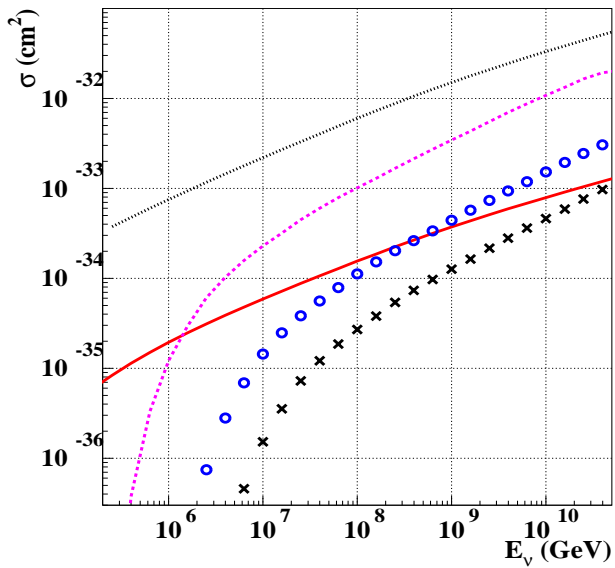,width=7.50cm,height=9.5cm,angle=0}
\caption{$\nu N$ cross sections vs. the energy of the incident neutrino.
The violet dashed, blue circled and black crossed lines correspond respectively
to 300, 600 and 900~GeV NLKPs. 
The top gray curve corresponds to the SM charged current interactions and the full
red one to the di-muon background. 
}
\end{center}
\label{fig:xs}
\end{figure}   

It is also interesting to compare the $\sigma_{\rm NLKP}$ to the NLSP
production
as obtained in Ref.~\cite{abcprl}. The NLKP production is significantly larger than 
the one for NLSPs, which translates into a larger number of events at the
detector, as we will see below. 

\section{NLKP Energy Loss}
\label{sec:dedx}
After production, the NLKPs lose energy due to ionization and 
radiation processes.
The average energy loss is given by~\cite{pdg}:
\beq
-\frac{dE}{dx} = a(E) + b(E)~E~,
\label{eq:eloss}
\eeq
where $a(E)$ represents the ionization losses, and $b(E)$ the contributions from
different radiation processes. The latter includes bremsstrahlung,
pair production and photo-nuclear interactions. There is also energy
loss due to weak
interactions, but this will only be of importance at very high
energy~\cite{weakloss}, and we will neglect it for the remaining of
this work. 

At the high energies where NLKPs can be produced, radiation losses
dominate over ionization. Among radiation processes, both pair production and
bremsstrahlung become less important for heavy particles. Although photonuclear
processes dominate tau lepton propagation losses~\cite{inadave,buga}, a mass
suppression will occur for leptons of much heavier masses \cite{abcprd,inarad}.

In order to determine the NLKP energy losses, we follow closely the calculations
done for NLSP propagation in Refs.~\cite{abcprd} and \cite{inarad}.  
Radiation losses dominate above a propagating energy of 1~TeV. 
Among them, pair production and bremsstrahlung are less
important for the NLKP when compared to photo-nuclear interactions, as
can be seen in Figure~3. Even so, the energy loss
due to photo-nuclear interactions is suppressed by the NLKP mass. 
As mentioned in Ref.~\cite{abcprl}
and shown explicitely in Refs.~\cite{abcprd} and \cite{inarad}, the important energy region 
for this process is the one at low photon virtuality $Q^2$. The reason is
that the structure function involved in the process is determined by a cross section
which is dominated by physics at low $Q^2 \simeq 1$~GeV$^2$. However, due to the
large NLKP mass, the minimum value for the photon virtuality will
be larger, therefore avoiding the effects of resonances and other nonperturbative processes
which occur at lower $Q^2$. This is  in contrast to the case of the
$\tau$ lepton, where the resonant region still dominates and results
in a much larger photo-nuclear energy loss.

Figure~3  shows the radiation loss term of eq.~(\ref{eq:eloss}) versus
neutrino energy for muons, taus and the 300~GeV NLKP. As expected, the photo-nuclear 
process
dominates the NLKP radiation loss. However, it is still quite suppressed due to the
NLKP heavy mass and the total energy loss is still considerably below the one for SM leptons.
Energy suppression will be enhanced for heavier NLKP mass.

\begin{figure}
\begin{center}
\epsfig{file=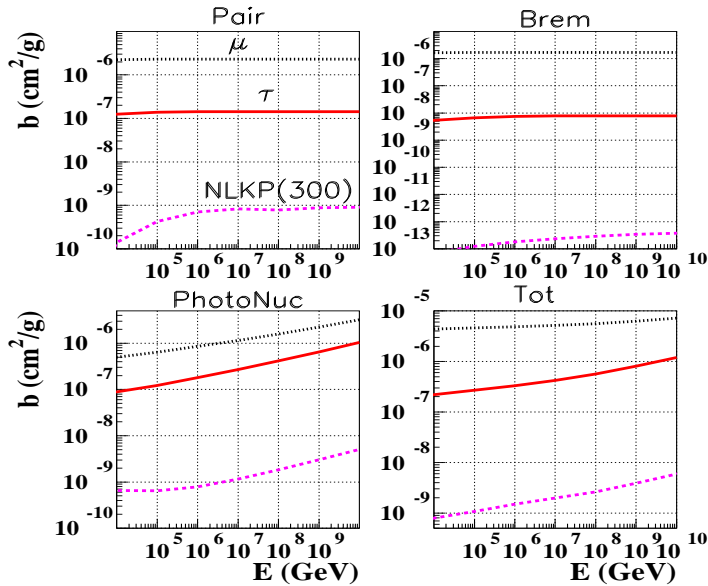,width=8.25cm,height=9.5cm,angle=0}
\caption{Radiation energy loss b(E) parameter due to pair production, bremsstrahlung and 
photonuclear processes for muon, tau and a 300~GeV NLKP. The plot labeled ``Tot'' is 
the sum of all contributions. The muon, tau and NLKP curves are as labeled in the first 
plot. Heavier NLKPs will have lower b(E) parameters.
}
\end{center}
\label{fig:eloss}
\end{figure}   
We then conclude that the NLKP energy loss is quite suppressed in
comparison with SM leptons. 
As we will see below, this means that its range
through the Earth is much larger, allowing for the detection 
of NLKPs that have been produced hundreds or even thousands of
kilometers from the detector.

\section{NLKP Signals and Rate in Neutrino Telescopes} 
\label{sec:signal}

\subsection{Neutrino Flux}

The NLKP event rate in neutrino telescopes depends on the incoming
neutrino flux. This is largely determined by the high
energy cosmic ray spectrum \cite{als}.
There are other potentially relevant sources of the neutrino flux,
such as atmospheric charm production~\cite{charmnus}. For the purpose
of this work we will neglect these other contributions, only
considering the flux of cosmic neutrinos, for which we use  two
alternative estimates:  
the work of Waxman and Bahcall (WB)~\cite{wb} and the
one of Manheim, Proterhoe and Rachen (MPR)~\cite{mpr}. The integrated number of
events resulting from the MPR limit is considerably larger than the WB.
We find our NLKP rates assuming each of these limits as our incoming neutrino
flux. All plots are produced assuming the WB limit as our neutrino flux.

Waxman and Bahcall fix the cosmic ray spectrum to a power law curve with spectral 
index $-2$.
The neutrino upper limit is deduced assuming that each nucleon will interact with
photons and produce a pion. The charged pions will then decay producing neutrinos.
Their argument requires that the sources are optically thin, which means that most of 
the protons escape the source without interacting. As a result, the neutrino
upper limit is given by
\be
\left(\frac{d\phi_\nu}{dE}\right)_{\rm WB} = \frac{(1-4) \times
10^{-8}}{E^2} {\rm GeV~ cm^{-2} s^{-1} ~sr{-1}}~,
\label{wblimit}
\ee
where the range in the coefficient depends on the cosmological
evolution of the sources. The evolution accounts for the source activity and redshift
energy loss due to the cosmological expansion.
We take the upper end as the neutrino flux incoming through the Earth.

On the other hand, instead of taking a fixed power law behaviour for all 
cosmic ray spectrum, Manheim, Proterhoe and Rachen determine the spectrum
at each energy directly from data. Here we consider the limit MPR obtain assuming
optically thin sources, although they also determine a limit for optically
thick sources (See comments about optically thick
sources in \cite{als}). Figure~4  shows both WB and MPR limits for
the muon plus anti-muon neutrino flux.

\begin{figure}
\begin{center}
\epsfig{file=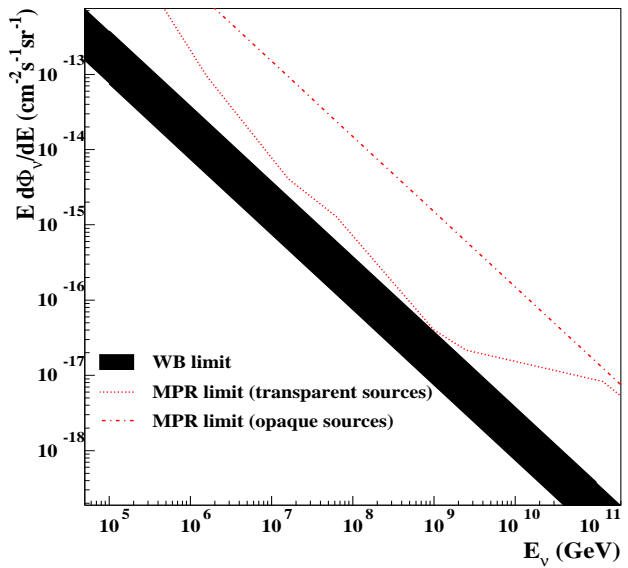,width=7.5cm,height=9.5cm,angle=0}
\caption{Upper bound on differential neutrino flux as calculated by
WB~\cite{wb} (shaded area) and MPR~\cite{mpr} (red lines). The WB limit
ranges from the limit with no cosmological evolution (upper edge) and with
cosmological evolution (lower edge). The MPR limit is shown for optically
thin sources (red dotted line) as used in this paper and for optically
thick sources (red dot-dashed line).}
\end{center}
\label{fig:nuflx}
\end{figure}   

As seen in Section~\ref{sec:xsecs}, the NLKP production is independent
of the initial neutrino flavor. For this reason we consider both electron and
muon neutrinos, and
neutrino mixing does not affect our results.

\subsection{NLKP Signals}
We now have all the ingredients to determine the NLKP rate at neutrino telescopes.
In order to understand the signal in detail, we performed a Monte Carlo
simulation generating approximately 30,000 events for each NLKP mass
(300, 600 and 900~GeV).

Once the incoming neutrino flux is determined, an interaction point is
randomly chosen based on the NLKP production probability. This results from
a convolution of the neutrino survival probability
with the probability of interacting and producing a NLKP.
The neutrino survival probability $P_S$ is given by $\exp(\int n dl)$, where $n$
is the Earth number density and $l$ is the distance the neutrino travels.
We use the Earth density profile as described in \cite{gqrs,eprof}.

The primary particles ($L^{(1)}_i$ and $Q^{(1)}$ produced in the 
neutrino interaction)
angular distribution at the CM is randomly determined 
based on the differential production cross section.
The center of mass (CM) angular distribution of the two NLKPs produced is 
assumed to be the same as the one between the two primary particles.
This is a good approximation \cite{abcprd} for events with energy well above
the production threshold where most of the event rate comes from. The events
close to the production threshold have a broader angular distribution. These would 
enhance the separation differences between signal and background and
therefore make our results conservative. The CM
angular distribution is then boosted to the laboratory frame.

\begin{figure}
%
\begin{center}
\epsfxsize=200pt \epsfbox{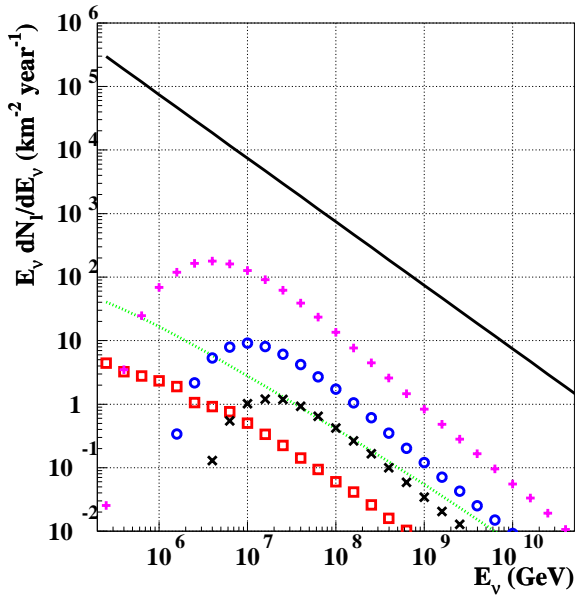}
\caption{NLKP pair event energy distribution per
km$^2$, per year, at the detector. Plus violet line corresponds
to 300 GeV; blue circled line to 600 GeV and crossed black line
to 900 GeV NLKP. For reference the neutrino flux at earth (full black
line); and the $\mu$ (dotted green line) and
the di-muon (squared red line) flux through the detector are also shown.
In all cases we make use of the WB limit for the neutrino flux. 
}
\label{fig:kken}
\end{center}
\end{figure}

Once the NLKPs are produced their propagation through the Earth is
simulated. Their energy loss -- which is mass dependent -- is taken into 
account. The NLKP energy distribution as a function of neutrino energy
is shown in Figure~\ref{fig:kken}. As can be seen, the 300 and 600 GeV NLKP 
event rate is much larger than the muon's for energies above NLKP production
threshold. The 900 GeV NLKP rate will be comparable with that for muons, but
still larger than the di-muon background rate. 

Although these are rather large rates, they do not directly translate
into observed NLKPs due to the fact that NLKPs are hard to identify.  
Neutrino telescopes measure their energy in two
ways\cite{als,ice3}: low energy events (below $\sim 100$~GeV) have their energy 
reconstructed from the track
length, whereas for the more energetic  ones the energy is
reconstructed 
from the amount of Cerenkov light deposited
in the photomultiplier tubes. Taking the Cerenkov radiation as 
proportional to the amount of deposited energy in the detector is a
good approximation for SM leptons. But the NLKPs lose a lot less energy than
SM leptons. Thus, if a NLKP track is assumed to be a SM lepton such as a
muon, it will be assigned a much lower energy as such.
For this reason and in order to compare event rates, 
the muon rate must be integrated from energies lower than the 
KK production threshold.

Table~\ref{tab:rate} shows the event rate per year per km$^2$ both for
the WB flux,  as well as for the MPR optically thin flux. The numbers
are clearly encouraging for km$^3$
neutrino telescopes. 
\begin{table}
\begin{center}
\caption{Number of events per km$^2$ per year for different NLKP masses and
neutrino fluxes at the Earth. The NLKP masses are 300, 600 and 900~GeV.
The number of NLKP events are given for energies
above threshold for production of a $L^{(1)}_\ell$ and  a $Q^{(1)}$ while
the muon rate for energies above 1000~GeV. The column $\mu^+\mu^-$ corresponds
to the di-muon background. No cuts were applied at this stage.}  
\begin{ruledtabular}
\begin{tabular}{l|ccccc}
 & $\mu$ & $\mu^+\mu^-$ & &$L^{(1)}_R\overline{L}^{(1)}_R$ & \\
~\\
& & & (300) & (600) & (900)  \\
\hline\\
WB & 552 & 30 & 489 & 21 & 3 \\
MPR & 39654 & 1914 & 1476 & 47 & 5 
\label{tab:rate}
\end{tabular}
\end{ruledtabular}
\end{center}
\end{table}
Two features will be important to distiguish the signal from
the background : the separation between the pair of particles
inside the detector; and -- for lower mass NLKPs -- a bump in the energy 
spectrum will appear. These features will be  discussed at the end of
this section.

\subsubsection{Di-muon Background}

Due to their large boost  
most NLKP pairs   go through the detector in
two well separated and approximately parallel tracks. Events well separated are produced far from the
detector and as the production angle between them is small the
tracks will be almost parallel. Therefore, single muons can be eliminated by a two
track requirement. The main remaining background are di-muons. These are produced
from charmed hadrons from the following chain :
\be
\nu N\to \mu^- H_c \to \mu^- ~\mu^+ ~H_x~\nu  ~,
\nonumber
\ee
where $H_c$ is a charm hadron produced from a muon neutrino CC interaction and
$H_x$ can be either a strange or non-strange hadron.

The cross section for charm production from a neutrino interaction was
calculated in
Ref.~\cite{abcprd}, as well as the di-muon energy loss, propagation and separation at the
detector. In what follows we reproduce these results, and compare  with the NLKP
signal.

\subsubsection{NLKPs Separation}

The separation between the NLKPs will be given by the distance traveled
times the angle ($\theta$) between the pair in the laboratory frame. As
the boost from CM to lab is large, $\theta$ is very small. However, this is
compensated by the production being far away from the detector. The 
production point being typically a few 1000 km from the detector and
$\theta \sim 10^{-4} - 10^{-5}$ the separation between the two NLKPs
will be a few tens to a few hundred meters.

On the other hand, di-muon events have to be produced close to the
detector, otherwise they lose
all their energy before arriving at it. For this reason 
their separation is typically smaller than the one for most of the signal events.

\begin{figure}
%
\begin{center}
\epsfig{file=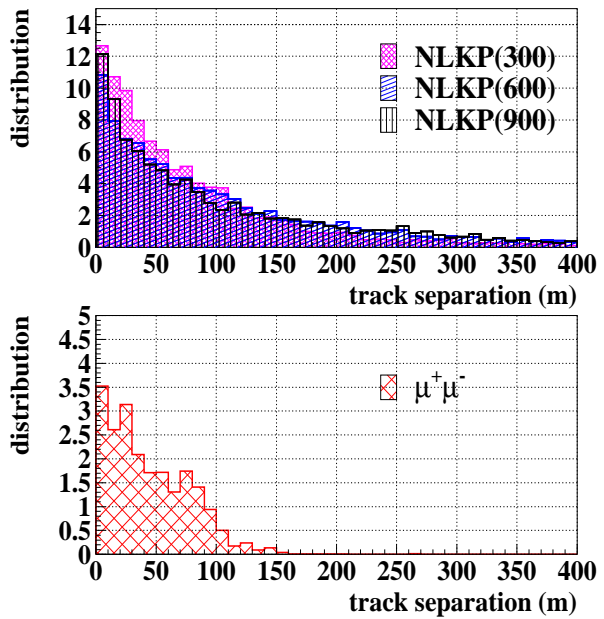,width=7.2cm,height=9cm,angle=0}
\caption{Track separation distribution between NLKP pair (top) and
for the di-muon background (bottom).  
}
\label{fig:sep}
\end{center}
\end{figure}

The separation distribution for each NLKP mass at the detector is shown in 
Figure~\ref{fig:sep}. The simulated detector is placed at the same depth as the IceCube
telescope \cite{ice3}. We also show the di-muon background separation
for comparison. While the dimuon separation is at most $\sim 100$~m, the pair
of NLKP can be more than 100 meters apart. 
For instance, for a 300~GeV NLKP, 52\% of the events
are more than 50~m apart and 28\% are more than 100~m apart. The di-muon
background has only 8\% with more than 50~m and 1.3\% with more than 100~m separation.
The 600 and 900~GeV NLKPs have both around 60\% of events with more than 50~m
separation and around 42\% with more than 100~m separation. 

In order to estimate the statistical significance of the separation cut,
we determine the $S/\sqrt{B}$ ratio, where $S$ and $B$ are respectively the number of 
signal and background events. We find that for the 300~GeV NLKP, a requirement
that the pair of NLKPs are at least 10 meters apart will yield a significance
of 85, ie, 436 of the 489 NLKPs will be more than 10 meters apart, while only 25
di-muons will have more than 10 meters separation. 
For the 600~GeV NLKP, a requirement of 86~m separation will allow a $5 \sigma$
significance in one year, with 9 signal events and 3 di-muons. For the 900~GeV, the
separation is harder, a $2 \sigma$ significance can be achieved in a year with a 
separation cut of 150~m, while a $5 \sigma$ significance needs 5 years to be achieved.

\subsubsection{The NLKP Bump}

\begin{figure}
\begin{center}
\epsfig{file=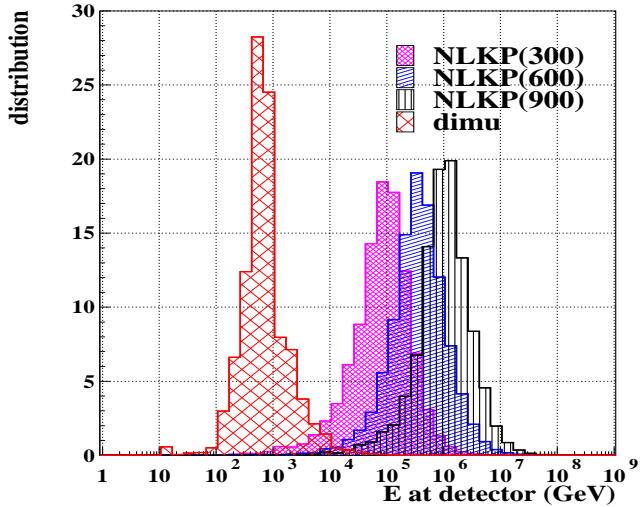,width=7.6cm,height=8.5cm,angle=0}
\caption{Arrival energy distribution 
of the NLKP at the 
detector for $m_{L^{(1)}_R}=300,~600 {\rm ~and~} 900$~GeV.
Also shown is the arrival distribution for the di-muon
background.}
\label{fig:eatd}
\end{center}
\end{figure}

Another feature of the NLKP signal comes from the fact that these particles
lose less energy than a SM lepton. This implies that NLKPs will have their energy
reconstructed as if they where lower energy leptons.
Figure~\ref{fig:eatd} shows both NLKP and di-muon simulated energy distribution as they
arrive at the detector. Although the NLKPs are more energetic than the di-muons, the
energy deposited in the PMTs will resemble lower energy muons and therefore they
have to be compared with them. However this will generate a sizeable
excess in the 
reconstructed energy spectrum, at least if the number of NLKP events
is large enough.
In order to understand how this feature will change the reconstructed energy
spectrum, we simulate the reconstructed energy by taking all NLKPs as muons.
This was done by determining the deposited energy in the detector and 
reconstructing this energy as if deposited by a muon. These events were
then added to the SM muon energy spectrum.
The consequence is that the high energy NLKPs will be reconstructed as
lower energy events that will end up as a bump around energies of TeVs.
\begin{figure}[t]
%
\begin{center}
\epsfig{file=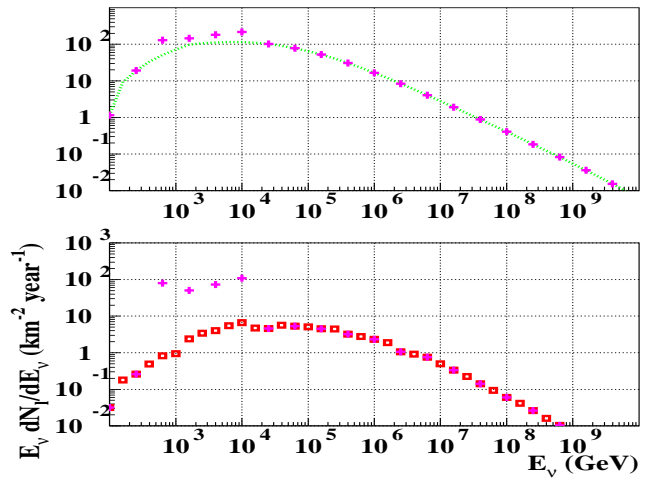,width=7.8cm,height=8cm,angle=0}
\caption{Top: Energy distribution of the muon flux through the detector (blue
circles) and the same flux plus the 300~GeV NLKP spectrum reconstructed as muons
(violet plus line). Bottom: Same as above but now using the di-muon flux through
the detector.
}
\label{fig:bump}
\end{center}
\end{figure}

Figure~\ref{fig:bump} shows the energy distribution of the muon flux through the
detector (top plot, blue circles) and the same distribution with the addition of 300 GeV
NLKPs reconstructed as muons. A visible ``crown'' with few events in each energy bin
in the 1 to 100 TeV region will clearly indicate the presence of KK particles. This
feature will be enhanced when the NLKPs are included in the di-muon energy spectrum
(bottom plot). When the signal is reconstructed as di-muons, a pronounced crown shows
up in the reconstructed energy spectrum. This feature is observable  for 
NLKPs in the lower mass range, 
since the rate of higher mass NLKPs would  not be large enough to
observably enhance the spectrum 
in the lower energy region. We expect this feature to be observable up
to NLKP masses of about $\sim 600~$GeV. Thus, for these lower mass
NLKPs there will be two distinct ways to separate the signal from the
main background.
\section{Conclusions}
\label{conc}
We have shown that in a UED scenario where the  NLKP is a the first KK
mode of a right handed charged lepton, neutrino telescopes such as
IceCube will be able to directly observe these $\ell^{(1)}_i$'s up to
masses of several hundred GeV, perhaps even $1~$TeV. This complements
hadron collider searches, where signals for this UED scenario would
consist of large missing energy, and perhaps one or two
highly-ionizing tracks. The similarity of the UED signals with the
analogous supersymmetric scenario, for instance with  gravitino dark
matter and a slepton NLSP, can make the identification  of the
underlying theory difficult. On the other hand, the event rate at 
neutrino telescopes coming from this UED scenario is considerably
higher than the one resulting from the supersymmetric case and studied
in Refs.~\cite{abcprl,abcprd}. 

We have made a detailed study of the background and the signal, and
shown that the track separation of NLKPs is a good discriminant with
respect to the di-muon background. In addition, 
for the case of smaller NLKPs masses, we have shown that the NLKP
signal results in a bump in the detected di-muon spectrum, since NLKPs
lose energy similarly to lower energy muons.  Combining this feature
with the characteristic track separation of the signal tracks should
enhance the statistical significance of a potential signal.

{\it Acknowledgments ---}
The authors thank Z.~Chacko for helpful discussions and for his
collaboration at early stages of this work.
 I.~A. and G.~B. acknowledge the support of the State of S\~{a}o Paulo 
Research Foundation (FAPESP) and the Brazilian National Counsel for 
Technological and Scientific Development (CNPq). C.~K. and B.~N. are 
supported by the NSF under grant PHY-0408954.


\begin{thebibliography}{99}

\bibitem{led} N.~Arkani-Hamed, S.~Dimopoulos and G.~R.~Dvali,
Phys.\ Lett.\ B {\bf 429}, 263 (1998); 
N.~Arkani-Hamed, S.~Dimopoulos and G.~R.~Dvali,
Phys.\ Rev.\ D {\bf 59}, 086004 (1999).
%

\bibitem{bogdan}
  T.~Appelquist, H-C. Cheng and B.~A.~Dobrescu,
  Phys.\ Rev.\ D\  {\bf 64}, 035002 (2001).


\bibitem{rs1}L.~Randall and R.~Sundrum,
  Phys.\ Rev.\ Lett.\  {\bf 83}, 3370 (1999).

\bibitem{cms1}
  H.~C.~Cheng, K.~T.~Matchev and M.~Schmaltz,
  Phys.\ Rev.\  D {\bf 66}, 036005 (2002)
  [arXiv:hep-ph/0204342].

\bibitem{FRT}
  J.~L.~Feng, A.~Rajaraman and F.~Takayama,
  Phys.\ Rev.\ Lett.\  {\bf 91}, 011302 (2003)
  [arXiv:hep-ph/0302215];
  J.~L.~Feng, A.~Rajaraman and F.~Takayama,
  Phys.\ Rev.\  D {\bf 68}, 063504 (2003)
  [arXiv:hep-ph/0306024];
  J.~L.~Feng, A.~Rajaraman and F.~Takayama,
  Phys.\ Rev.\  D {\bf 68}, 085018 (2003)
  [arXiv:hep-ph/0307375].

\bibitem{MSSY}
  S.~Matsumoto, J.~Sato, M.~Senami and M.~Yamanaka,
  Phys.\ Lett.\  B {\bf 647}, 466 (2007)
  [arXiv:hep-ph/0607331];
  S.~Matsumoto, J.~Sato, M.~Senami and M.~Yamanaka,
  Phys.\ Rev.\  D {\bf 76}, 043528 (2007)
  [arXiv:0705.0934 [hep-ph]].

\bibitem{muedphase}
J.~A.~R.~Cembranos, J.~L.~Feng and L.~E.~Strigari,
  Phys.\ Rev.\  D {\bf 75}, 036004 (2007)
  [arXiv:hep-ph/0612157].

\bibitem{sixd}G.~Burdman, B.~A.~Dobrescu and E.~Ponton,
  JHEP {\bf 0602}, 033 (2006)
  [arXiv:hep-ph/0506334]; G.~Burdman, B.~A.~Dobrescu and E.~Ponton,
  Phys.\ Rev.\  D {\bf 74}, 075008 (2006)
  [arXiv:hep-ph/0601186].

\bibitem{abcprl}
  I.~Albuquerque, G.~Burdman and Z.~Chacko,
  Phys.\ Rev.\ Lett.\  {\bf 92}, 221802 (2004).

\bibitem{Bi}
  X.~J.~Bi, J.~X.~Wang, C.~Zhang and X.~m.~Zhang,
  Phys.\ Rev.\  D {\bf 70}, 123512 (2004)
  [arXiv:hep-ph/0404263].

\bibitem{AKR}
  M.~Ahlers, J.~Kersten and A.~Ringwald,
  JCAP {\bf 0607}, 005 (2006)
  [arXiv:hep-ph/0604188].

\bibitem{abcprd}
  I.~Albuquerque, G.~Burdman and Z.~Chacko,
  Phys.\ Rev.\ D\  {\bf 75}, 035006 (2007).

\bibitem{AIMM}
  M.~Ahlers, J.~I.~Illana, M.~Masip and D.~Meloni,
  JCAP {\bf 0708}, 008 (2007)
  [arXiv:0705.3782 [hep-ph]];
  M.~H.~Reno, I.~Sarcevic and J.~Uscinski,
  Phys.\ Rev.\  D {\bf 76}, 125030 (2007)
  [arXiv:0710.4954 [hep-ph]];
  S.~Ando, J.~F.~Beacom, S.~Profumo and D.~Rainwater,
  arXiv:0711.2908 [hep-ph].

\bibitem{pdg} W.~M.~Yao {\it et al.}  [Particle Data Group],
J.\ Phys.\ G {\bf 33}, 1 (2006).

\bibitem{inadave} S.~I.~Dutta, M.~H.~Reno, I.~Sarcevic and D.~Seckel,
Phys.\ Rev.\ D {\bf 63}, 094020 (2001).

\bibitem{buga} E.~V.~Bugaev and Y.~V.~Shlepin,
Phys.\ Rev.\ D {\bf 67}, 034027 (2003).

\bibitem{inarad} M.~H.~Reno, I.~Sarcevic and S.~Su,
Astropart.\ Phys.\  {\bf 24}, 107 (2005).

\bibitem{weakloss} Y.~Huang, M.~H.~Reno, I.~Sarcevic and J.~Uscinski,
  Phys.\ Rev.\  D {\bf 74}, 115009 (2006)

\bibitem{als} For a detailed discussion of neutrino event rates see
I.~Albuquerque, J.~Lamoureux and G.~F.~Smoot,
Astrophys.\ J.\ Suppl.\  {\bf 141}, 195 (2002).

\bibitem{charmnus} S.~Ando, J.~F.~Beacom, S.~Profumo and D.~Rainwater,
  arXiv:0711.2908 [hep-ph].

\bibitem{wb}
E.~Waxman and J.~N.~Bahcall,
Phys.\ Rev.\ D {\bf 59}, 023002 (1999); 
J.~N.~Bahcall and E.~Waxman,
Phys.\ Rev.\ D {\bf 64}, 023002 (2001).

\bibitem{mpr}
K.~Mannheim, R.~J.~Protheroe and J.~P.~Rachen,
Phys.\ Rev.\ D {\bf 63}, 023003 (2001)

\bibitem{gqrs}
R.~Gandhi, C.~Quigg, M.~H.~Reno and I.~Sarcevic,
Astropart.\ Phys.\  {\bf 5}, 81 (1996); 
R.~Gandhi, C.~Quigg, M.~H.~Reno and I.~Sarcevic,
Phys.\ Rev.\ D {\bf 58}, 093009 (1998). 

\bibitem{eprof} A.~Dziewonski, in Encyclopedia of Solid Earth Geophysics,
ed. J.D.E. Van Nostrand (New York: Reinhold), 331 (1989).

\bibitem{ice3}
J.~Ahrens {\it et al.}  [The IceCube Collaboration],
Nucl.\ Phys.\ Proc.\ Suppl.\  {\bf 118}, 388 (2003).

\end{thebibliography}
\end{document}